\documentclass[useAMS]{mn2e}
\usepackage{epsf}
\usepackage{german}
\usepackage[dvips]{graphics}

\begin{document}
\title{Analysing Large Scale Structure: I. Weighted Scaling Indices
and Constrained Randomisation}

\author[C. R\"ath et al.]
  {Christoph R\"ath \thanks{E-mail: cwr@mpe.mpg.de}, Wolfram Bunk,
   Markus B. Huber, Gregor E. Morfill,
   \newauthor
    J\"org Retzlaff and Peter Schuecker\\
    Centre for Interdisciplinary Plasma Sciences (CIPS)/ \\
    Max-Planck-Institut f\"ur extraterrestrische Physik (MPE),
    Garching, Germany
  }

\date{Accepted 2002 July 2
      Received 2002 July 2;
      in original form 2002 January 31}

\pubyear{2002} \volume{000} \pagerange{1} \twocolumn

\maketitle

\label{firstpage}

\begin{abstract}

The method of constrained randomisation, which was originally
developed in the field of time series analysis for testing for
nonlinearities, is extended to the case of three-dimensional
point distributions as they are typical in the analysis of the large
scale structure of galaxy distributions in the universe.\\
With this technique it is possible to generate for a given
data set so-called surrogate data sets which have the same linear
properties as the original data whereas higher order or
nonlinear correlations are not preserved. The analysis of the
original and surrogate data sets with measures, which are
sensitive to nonlinearities, yields valuable information about
the existence of nonlinear correlations in the data.
On the other hand one can test whether given statistical
measures are able to account for higher order or nonlinear
correlations by applying them to original and surrogate data sets.\\
We demonstrate how to generate surrogate data sets from a given point
distribution, which have the same linear properties
(power spectrum) as well as the same density amplitude distribution
but different morphological features.\\
We propose weighted scaling indices, which measure the
local scaling properties of a point set, as a nonlinear statistical
measure to quantify local morphological elements in
large scale structure. Using surrogates is is shown that the data
sets with the same 2-point correlation functions have
slightly different void probability functions and especially a
different set of weighted scaling indices.\\
Thus a refined analysis of the large scale structure becomes
possible by calculating local scaling properties whereby
the method of constrained randomisation yields a vital tool for
testing the performance of statistical measures in
terms of sensitivity to different topological features
and discriminative power.

\end{abstract}


\section{Introduction}
One of the important issues in cosmology today is characterising the
nature of the large scale structure in the spatial distribution of galaxies
as revealed by observations. Statistical measures provide important tools
for the quantitative characterisation of the morphology of the galaxy distribution
and for the comparison of the various cosmological models with observations.
Among the first and still most frequently used measures are the 2-point
correlation function (e.g. Peebles 1980 and references therein;
Norberg et al. 2001) and the power spectrum (e.g. Szalay et al. 2001;
Tegmark et al. 2001; Schuecker et al. 2001)
which have the advantage of being directly
related to simulations for different cosmological models.
However, they are linear measures which cannot provide any information
about higher order or nonlinear correlations in the data
set. Nowadays the large surveys like the SDSS (York et al 2000)
or 2dF (Colless et al. 2001) yield excellent observations from
galaxy distributions consisting of up to one
million galaxies with which it becomes possible to identify higher
order correlations. Therefore it is necessary to develop statistical
descriptors, which go beyond the 2-point correlation function.\\
Many measures which go beyond the 2-point correlation function
have already been studied in detail. The correlation
analysis of the data sets has very early been extended to
higher order correlation functions (e.g. 3-point correlation function
(Groth \& Peebles 1977),  4-point correlation function
(Fry \& Peebles 1978), up to 8-point correlation function
(Meiskin,  Szapudi and \& Szalay 1992))
and are now applied to the newest available data
sets (Szapudi et al. 2002).
Analysis in the Fourier space have
involved the calculation of eigenvectors of the sample correlation
matrix (e.g. Vogeley et al. 1996) and of the
bispectrum (Mataresse, Verde \& Heavens 1997;
Verde et al. 1998; Scoccimarro et al. 2001). More recently, also the correlations
between Fourier phases have been
quantified by calculating entropies (Chiang \& Coles 2000, Chiang 2001),
which measure the amount of non-gaussian signatures in the
spatial patterns of a density field.
Other measures have been developed in order to characterise the
topology of the large scale structure. Among the first measures of this
kind introduced in cosmology has been the void probability function
(e.g. White 1979; Ghigna et al. 1994), which can be expressed
by a sum over all n-point correlation functions.
Another well-known measure is the genus curve of the
density contrast (Weinberg, Gott \& Mellott 1987), which has only recently
been applied to the 2dF galaxy redshift survey data set
(Hoyle, Vogeley \& Gott 2002).
Both the void probability function and the genus curve can be regarded
as special cases of the Minkowsky functionals which also have
extensively been used in the analysis of the galaxy distributions
(e.g. Mecke et al. 1994; Kerscher et al. 1997; Bharadwaj et al. 2000).
The concepts derived in the field of non-linear dynamics have been applied
to large scale structure analysis by calculation e.g.
the multifractal dimension spectrum (e.g. Borgani 1995 and references therein;
Pan \& Coles 2000).
One common feature of all these measures is that they analyse the
data set as a whole and therefore focus on
the {\it global} aspects of matter distribution.\\
In the field of image analysis various statistical
methods for the morphological
and textural description of given structures have been developed, too
(for an overview see e.g. Tuceryan \& Jain 1993 and references therein).
It has been shown that in the context of (human) texture analysis it is
crucial to consider both global and {\it local} aspects
of given structures in order to perform an effective structure
characterisation (Sagi \& Julesz 1985; Jain \& Farrokhnia 1991)
leading e.g. to texture detection and discrimination.
Furthermore it has been pointed out (Julesz 1981, 1991) that nonlinear and
local data processing steps play a crucial role in the detection
and discrimination of textural features. It has been shown
that nonlinear local filters (so-called scaling indices)
which measure the local scaling properties of point sets
are well suited to accomplish feature and texture detections tasks in
image processing (R\"ath \& Morfill 1997; Jamitzky et al. 2001).
The general approach for estimating these measures, which is closely
related to the formalism of the multifractal dimension spectrum,
makes them ideal candidates for describing
the local structural features in galaxy distributions, too.
In this paper we propose a modified version of the scaling index
formalism ('weighted scaling indices') as a local nonlinear statistical
measure for analysing the large scale structure in the universe.\\
For the assessment of the different statistical measures it is of vital
interest to have detailed knowledge about the performance of the
different measures in terms of sensitivity to certain
morphological features or in terms of discrimination power.
In the analysis of nonlinear time series (Theiler et al. 1992;
Schreiber \& Schmitz 1996; Schreiber \& Schmitz 1997; Schreiber 1998)
the technique of constrained randomisation, that allows a test for
weak nonlinearities in time series, has been developed.
Applying this method to a given data set one obtains an
ensemble of randomised versions of the original data set
(so-called surrogate data), in which some previously defined statistical
constraints are maintained while all other
properties are subject to randomisation.
Using a different reasoning, one can also use
this method in order to test whether given statistical measures
are able to account for higher order or nonlinear correlations or
special morphological features in the data applying the measures
to be tested to both the original data and the surrogates and
comparing their discriminative power.
In this work we extend known techniques for generating surrogates to the
case of three-dimensional point distributions as they are typical in the
analysis of the large scale structure. We calculate several linear and
nonlinear measures for the data and surrogates and evaluate them in terms
of sensitivity and discriminative power.\\
The outline of the paper is as follows: In the next Section the properties
of the simulated data set are briefly described. In Section 3 we introduce
the statistical measures we used in our study.
Whilst the well-known measures used for references are only briefly reviewed,
the phase entropy and the concept of weighted scaling indices are described
in more detail. In Section 4 the results of our calculations are shown.
Section 5 contains the main conclusions and gives an outlook
for future work.

\section{The Data Set}
The method of constrained randomisation
is developed and tested using
$N$-body simulation data of a realistic cosmological model.
The simulation was performed using a
A$\rm P^{3}$M code (Couchman 1991) with $128^3$ particles
in a $(100\,h^{-1}\mathrm{Mpc})^3$ box on a $128^3$ grid with
the softening parameter $\epsilon=0.03$
(spatial force resolution $\approx 23.4\,h^{-1}\mathrm{kpc}$).
An OCDM model was simulated with total matter density parameter
$\Omega_0 = 0.35$, Hubble parameter $h=0.7$, no cosmological constant
($\Omega_{\Lambda} = 0$), normalisation amplitude $\sigma_8 = 0.78$, and
baryonic matter density parameter $\Omega_\mathrm{b}h^2 = 0.0125$.
For the transfer function the parametrization of Bardeen et al.\ (1986)
with the scaling proposed by Sugiyama (1995) was used.
The normalisation is compatible with the
abundance of clusters of galaxies in the universe (e.\,g.\ Eke et al.\ 1996).
The simulation was started at redshift $z=48$ (initial pertubations
imposed on the glass-like initial load using the
Zel'dovich approximation) and stopped after $1000$ time steps.
The code integrates the equations of particle motion using $p=a^{3/2}$
as a time variable, where $a$ is the scale factor.
From the $128^3$ simulated particles $5 \times 10^4$ particles
were randomly chosen.
This subset of the simulated and surrogate particles
and its respective point distribution
in the real space represents the basic data set for all
investigations in this study.
Similar investigations in the redshift space
are deferred to future work. The principle
outline of the following is independent of the
actually chosen configuration space.

\section{Statistical Measures}
In this section we introduce the statistical measures
used to analyse the point distributions in
this study. First we briefly summarise conventional measures,
namely the power spectrum, the 2-point correlation
function and the void probability function.
Then we describe in more detail
the phase entropy and especially the weighted
scaling indices, which are not so familiar in
this context.

\subsection{Conventional measures}
The density contrast $\delta_{\vec{r}}$ is given by
\begin{equation}
  \delta_{\vec{r}} = \frac{\rho_{\vec{r}}- < \rho > }{< \rho >} \;,
\end{equation}
where $\rho_{\vec{r}}$ denotes the point density at point $\vec{r}$. For
the determination of the power spectrum $P(k)$ of the density contrast
the standard estimator which takes into account the effect of the
discrete sampling of a point process is used:
\begin{equation}
  P(k) = (< | \delta_{\vec{k}} |^2> - \frac{1}{N_p}) \cdot L^3 \;.
\end{equation}
$\vec{k}$ is the wavenumber, $\delta_{\vec{k}}$ the Fourier transformed density contrast,
$N_p$ the number of points and $L$ denotes the size of the (cubic) volume.
The spatial 2-point correlation function, $\xi(r)$, is closely related to the power spectrum
but estimated in the configuration space. It can be defined through the joint probability
\begin{equation}
  \delta^2 P = \rho^2 \delta V_1 \delta V_2 (1 + \xi(r_{12}))
\end{equation}
of finding an object in the volume element $\delta V_1$ and another one in $\delta V_2$
at separation $r_{12}$ ($\rho$ being the mean point density).
The spatial 2-point correlation function $\xi(r)$ is a measure for the departure
from poissonian statistics.
Following the proposition of Hamilton (1993) we use in all our calculations the
estimator
\begin{equation}
  1 + \xi(r) = \frac{DD \cdot RR}{DR \cdot DR} \;,
\end{equation}
where $DD$ denotes the number of distinct pairs in the data, $RR$ denotes
the number of distinct pairs in the random distribution, and $DR$ the
number of cross pairs.\\
As a measure which, in general, depends on all higher order correlation
functions (White 1979) we calculate the void probability function (VPF)
$P_0(r)$. In order to estimate the VPF, we sample the point sets with random
spheres of different radii $r$. Centers are taken to be at distances greater
than $r$ from the boundaries of the point distribution. We take $N= 10000$ such spheres
and estimate the probability of finding an empty sphere.

\subsection{Phase Entropy}

Due to the nonlinear evolution of the large scale structure of
the universe the Fourier modes do not evolve independently - they
are coupled. In the highly non-gaussian regime the phases becomes
non-randomly distributed, containing information about
the underlying shape of the density distribution. Therefore the
analysis of the complete set of Fourier phases yields
statistical measures which may quantify the non-gaussian features
in the density field. Following the ideas of
Polygiannakis \& Moussas (1995) it has been proposed
(Chiang \& Coles 2000; Coles \& Chiang 2000; Chiang 2001)
to quantify the information contained in the phases by the
entropy $S(D_{k_i})$ of the phase gradients $D_{k_i}$,
\begin{equation}
  S = - \int_{-\pi}^{\pi} f(D_{k_i}) \ln (f(D_{k_i})) dD_{k_i} \;,
\end{equation}
where $f(D_{k_i})$ is the probabiltity function for $D_{k_i}$. $S$ becomes
maximal ($S = \ln(2\pi)$) if the density field is gaussian.
Non-gaussianity yields lower values for S.
In our discrete case the expression for the phase entropy becomes
\begin{equation}
  S_i = - \sum_{j=1}^{m} f(D_{k_i}(j))  \ln (f(D_{k_i}(j))) \delta D_{k_i}\;,
\end{equation}
where $D_{k_i} = \phi_{k_i}(k_i(j+1)) - \phi_{k_i}(k_i(j)), i=x,y,z$ denotes the directional
phase difference between adjacent phases. For our simulations with a resolution
of $128$ bins in each direction we have $1 \le k_i \le 64$ the upper limit being the
Nyquist frequency of the simulations. We calculate $S_i$ for each direction separately
and use the mean entropy $<S> = (S_x+S_y+S_z)/3$ as an estimator for the
information contained in the phases.

\subsection{Weighted scaling indices}
The basic concepts of this formalism have been developed in the context of the analysis of the nonlinear
system where it has been shown that global as well as local scaling properties of the phase space
representation of the system yield useful measures which characterise the underlying dynamics of the system
(for a comprehensive review see e.g. Paladin \& Vulpiani 1987). Based on these ideas we propose a modified
version of the estimation of {\it local} scaling properties of a point set - called
weighted scaling indices (WSI) - and apply this method in order to charaterise different
structural features in (simulated) particle distributions.
Consider a set of $N$ points $P=\{\vec{p_i}\}, i=1,\ldots,N$.
For each point the local weighted cumulative point distribution $\rho$ is calculated.
In general form this can be written as
\begin{equation}
  \rho(\vec{p_i},r) = \sum_{j=1}^{N} s_r (d(\vec{p_i},\vec{p_j})) \;,
\end{equation}
where $s_r(\bullet)$ denotes a shaping function depending on the scale parameter $r$
and $d(\bullet)$ a distance measure.\\
The weighted scaling indices $\alpha(\vec{p_i},r)$ are obtained by calculating
the logarithmic derivative of $\rho(\vec{p_i},r)$ with respect to $r$,
\begin{equation}
  \alpha(\vec{p_i},r) = \frac{\partial \log \rho(\vec{p_i},r)}{\partial \log r}
                      = \frac{r}{\rho}\frac{\partial}{\partial r} \rho(\vec{p_i},r) \;.
\end{equation}
In principle any differentiable shaping function and any distance measure can
be used for calculating $\alpha$. In the following we use the euclidean norm as
distance measure and a set of gaussian shaping function. So the expression for $\rho$
simplifies to
\begin{equation}
  \rho(\vec{p_i},r) = \sum_{j=1}^{N} e^{-(\frac{d_{ij}}{r})^q} \;,
                       d_{ij} = \| \vec{p_i} - \vec{p_j} \| \;.
\end{equation}
The exponent $q$ controls the weighting of the points according to their
distance to the point for which $\alpha$ is calculated. For small values
of $q$ points in a broad region around $\vec{p_i}$ significantly
contribute to the weighted local density $\rho(\vec{p_i},r)$.
With increasing values for $q$ the shaping function becomes more and more
a steplike function counting all points with $d_{ij} < r$ and neglecting
all points with $d_{ij} > r$. In this study we calculate $\alpha$ for the case
$q=2$. Using the definition in (9) yields for the weighted scaling indices
\begin{equation}
  \alpha(\vec{p_i},r) = \frac{\sum_{j=1}^{N} q (\frac{d_{ij}}{r})^q e^{-(\frac{d_{ij}}{r})^q}}
                             {\sum_{j=1}^{N} e^{-(\frac{d_{ij}}{r})^q}} \;.
\end{equation}
Structural components of a point distribution are characterised by the calculated
value of $\alpha$ of the points belonging to a certain kind of structure.
For example, points in a cluster-like structure have $\alpha \approx 0$
and points forming filamentary structures have $\alpha \approx 1$. Sheet-like structures
are characterised by $\alpha \approx 2$ of the points belonging to them. A uniform distribution
of points yields $\alpha \approx 3$ which is equal to the dimension of the configuration space.
Points in underdense regions in the vicinity of point-like structures, filaments or walls
have $\alpha > 3$.\\
The parameter $r$ determines the length scale on which the structures are analysed.
Obviously the value of $\alpha$ strongly depends on the choice of $r$. If $r$
approaches zero, each point 'sees' no neighbours due to the sharp decrease
of the shaping function with increasing distance $\| \vec{p_i} - \vec{p_j} \| $.
So each point forms a pointlike structure ($\alpha =0$) with itself as only member.
If $r$ has the same length scale as the structures to be analysed, one
obtains the full spectrum of $\alpha$ values belonging to different structural elements
(provided they are realized in the point distribution). If $r$ is further increased
the differences of the structural elements become less pronounced whereas
edge effects begin to play an important role. Thus the frequency distribution
narrows and shifts to lower values of $\alpha$.\\
The scaling indices for the whole point set under study form the frequency
distribution $N(\alpha)$
\begin{equation}
  N(\alpha) d\alpha = \#(\alpha \in [ \alpha,\alpha+d\alpha [)
\end{equation}
or equivalently the the probability distribution
\begin{equation}
  P(\alpha) d\alpha = Prob(\alpha \in [ \alpha,\alpha+d\alpha [)
\end{equation}
This representation of the point distribution can be regarded as a
structural decomposition of the point set where
the points are differentiated according to the local
morphological features of the structure elements to which they belong to.
Thus the spectrum reveals the structural content of a point set under study.

\section{Constrained Randomisation}
In the method of constrained randomisation an ensemble of surrogate
data sets are generated which share given properties
of the observed point distribution.\\
In our case we want the surrogate data sets to have the same power spectrum
in Fourier space and the same amplitude distribution of the point
density in configuration space as the original data set.
A sophisticated approach, which fulfills these requirements, is the method
of iteratively refined surrogates (Schreiber \& Schmitz 1996). In this section
we propose a three-dimensional extension of the method.\\
The algorithm consists of an iteration scheme. Before the iteration begins
two quantities have to be calculated:\\
1) A copy $\eta(\vec{r}) = \mbox{rank}(\rho(\vec{r}))$ of the original, coarse grained
three-dimensional discrete density field $\rho(\vec{r})$, which is sorted
by magnitude in ascending order, is computed.\\
2) The absolute values of the amplitudes of the Fourier
transform $\rho(\vec{k})$ of  $\rho(\vec{r})$,
\begin{equation}
   \left| \rho(\vec{k}) \right| =
   \left| \frac{1}{N_{bins}^3} \sum_{i,j,k=0}^{N_{bins}-1}
           \rho(\vec{r}_{ijk}) e^{-2 \pi i \vec{k} \vec{r}/N_{bins}} \right|
\end{equation}
are calculated as well. Both quantities $\eta(\vec{r})$ and
$\left| \rho(\vec{k}) \right|$ are stored for later use.\\
The starting point for the iteration is a random
shuffle $\rho_0(\vec{r})$ of tha data. Each iteration
consists of two consecutive calculations:\\
First $\rho_0(\vec{r})$ is brought to the desired sample
power spectrum. This is achieved by using a crude 'filter' in the
Fourier domain: The Fourier amplitudes are simply {\it replaced} by the
desired ones.\\
For this the Fourier transforma of $\rho_n(\vec{r})$ is taken:
\begin{equation}
 \rho_n(\vec{k}) = \frac{1}{N_{bins}^3} \sum_{i,j,k=0}^{N_{bins}-1}
           \rho_n(\vec{r}) e^{-2 \pi i \vec{k} \vec{r}/N_{bins}} \;.
\end{equation}
In the inverse Fourier transformation the actual amplitudes are
replaced by the desired ones and the phases defined
by $ \tan{\psi_n(\vec{k})} = Im(\rho_n(\vec{k}))/Re(\rho_n(\vec{k}))$
are kept:
\begin{equation}
 s(\vec{r}) = \frac{1}{N_{bins}^3} \sum_{k_x,k_y,k_z=0}^{N_{bins}-1}
           e^{i \psi_n(\vec{k})} \left| \rho(\vec{k}) \right|
	   e^{-2 \pi i \vec{k} \vec{r}/N_{bins}} \;.
\end{equation}
Thus this step enforces the correct power spectrum but usually the
distribution of the amplitudes in the state space will be modified.\\
Second a rank ordering of the resulting data set $s(\vec{r})$ is
performed in order to adjust the spectrum of amplitudes. The amplitudes
$\rho_{n+1}(\vec{r})$ are obtained by replacing the values of
$s(\vec{r})$ with those stored in $\eta(\vec{r})$ according to
their rank:
\begin{equation}
 \rho_{n+1}(\vec{r}) = \eta(\mbox{rank}(s(\vec{r}))) \;.
\end{equation}
It is clear that the Fourier spectrum of $\rho_{n+1}(\vec{r})$,
$\rho_{n+1}(\vec{k})$, will differ from $\rho_{n}(\vec{k})$.
Thus the two steps have to be repeated several times until the power spectrum
of the surrogate data matches that of the original data within
a desired accuracy. It can be understood heuristically that
the iteration scheme is attracted to a fixed
point $\rho_{n+1}(\vec{r})=\rho_{n}(\vec{r})$ for large $n$. Since
the minimal possible change equals the the smallest nonzero difference
in $\eta(\vec{r})$ and is therefore finite for finite $N_{bins}$, the fixed
point is reached after a finite number of iterations.
The final accuracy that can be reached depends on the size
and structure of the data and is generally sufficient
for testing statistical measures for large scale structure.
\begin{figure*}
  \begin{center}
   \begin{minipage}{8.5cm}
     \epsfxsize=8.5cm
     \epsffile{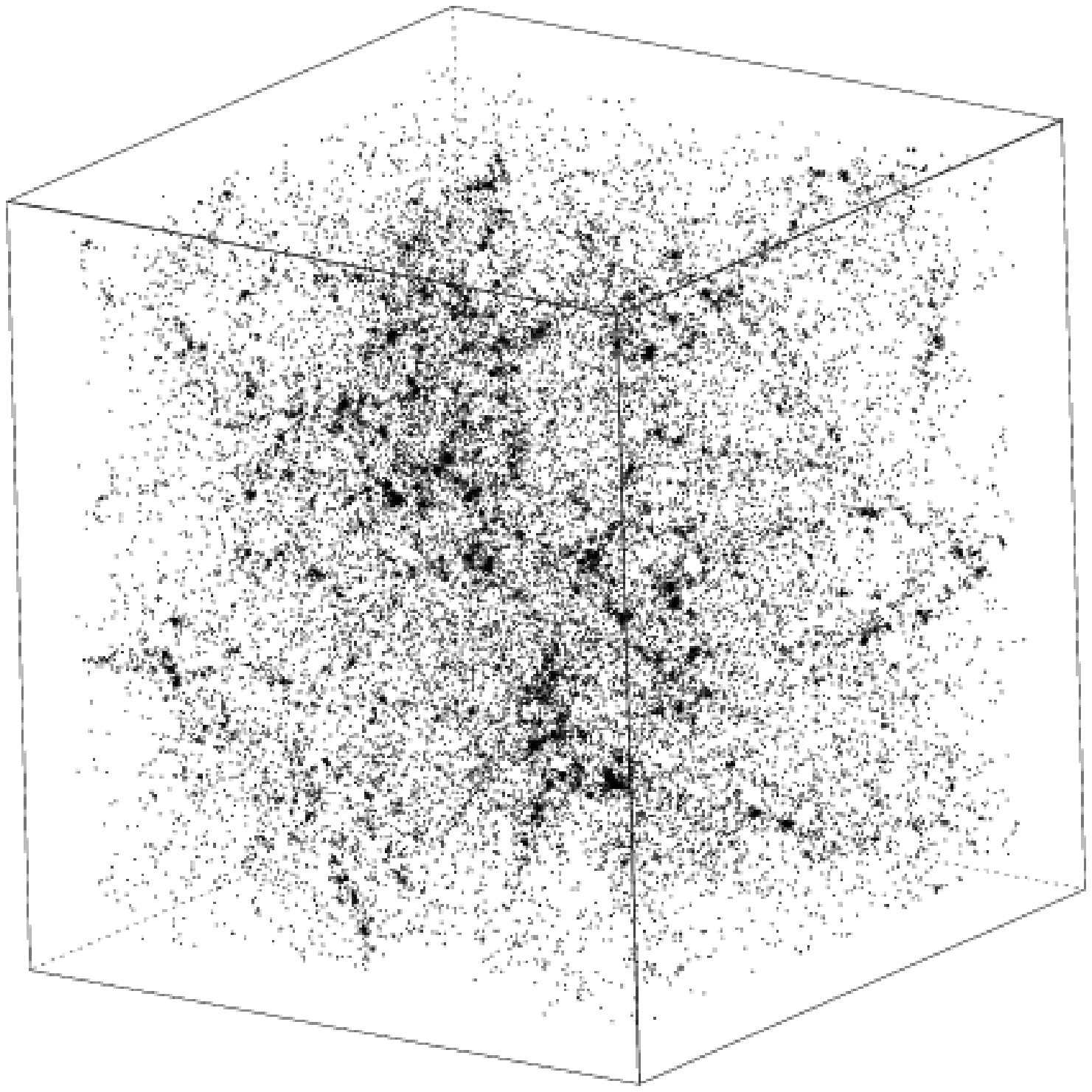}
    \end{minipage}
    \begin{minipage}{8.5cm}
    \epsfxsize=8.5cm
     \epsffile{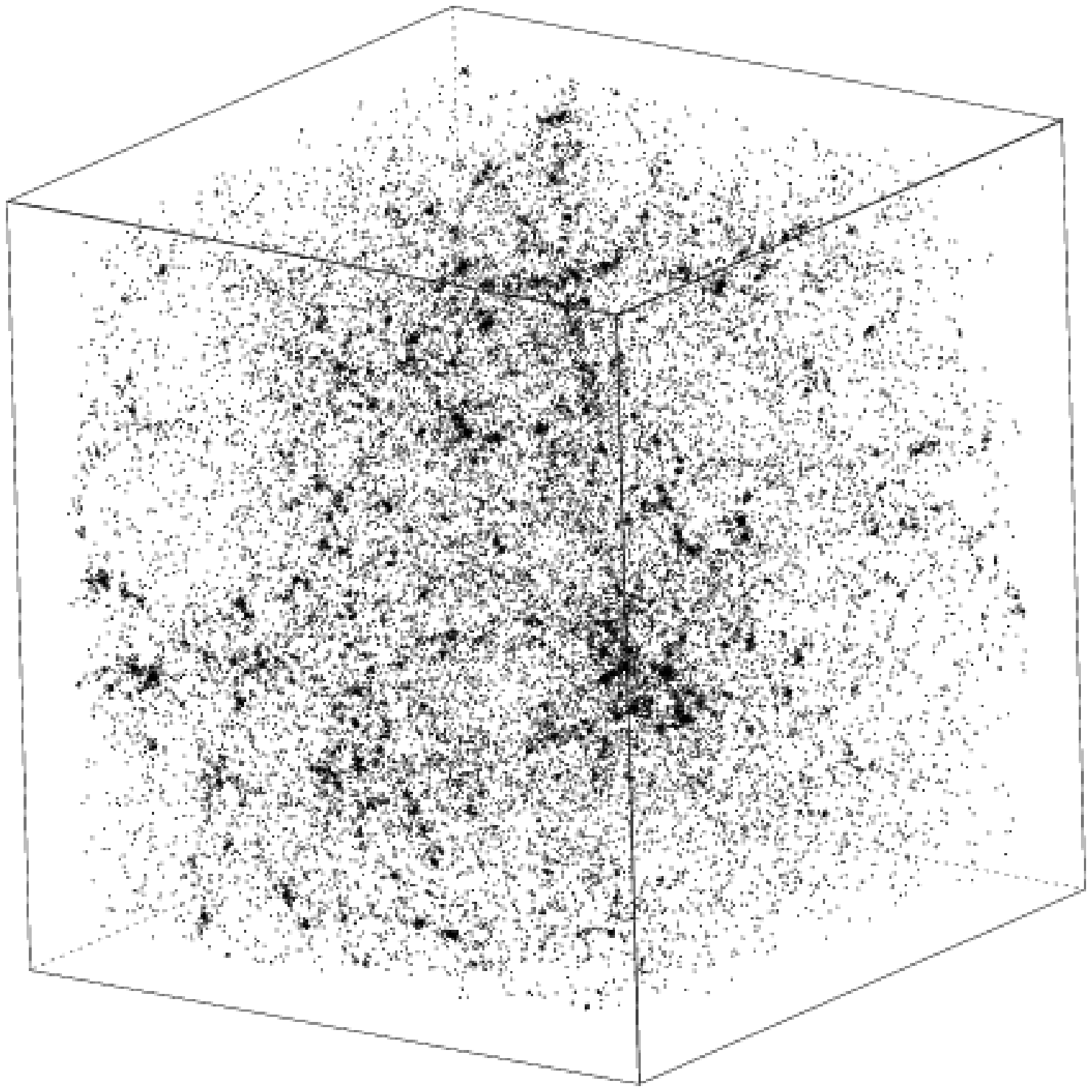}
    \end{minipage}

   \begin{minipage}{8.5cm}
     \epsfxsize=8.5cm
     \epsffile{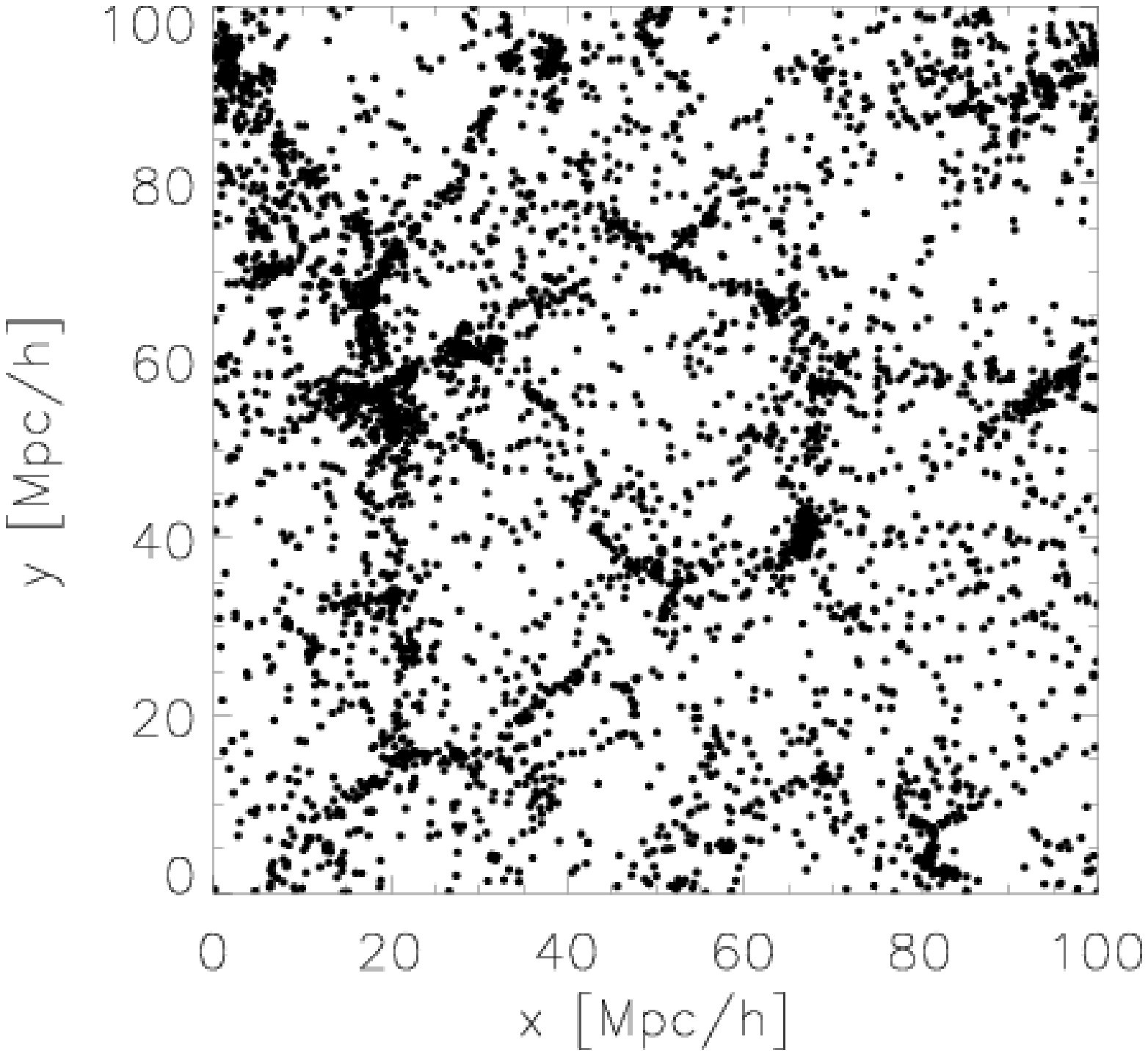}
    \end{minipage}
    \begin{minipage}{8.5cm}
    \epsfxsize=8.5cm
     \epsffile{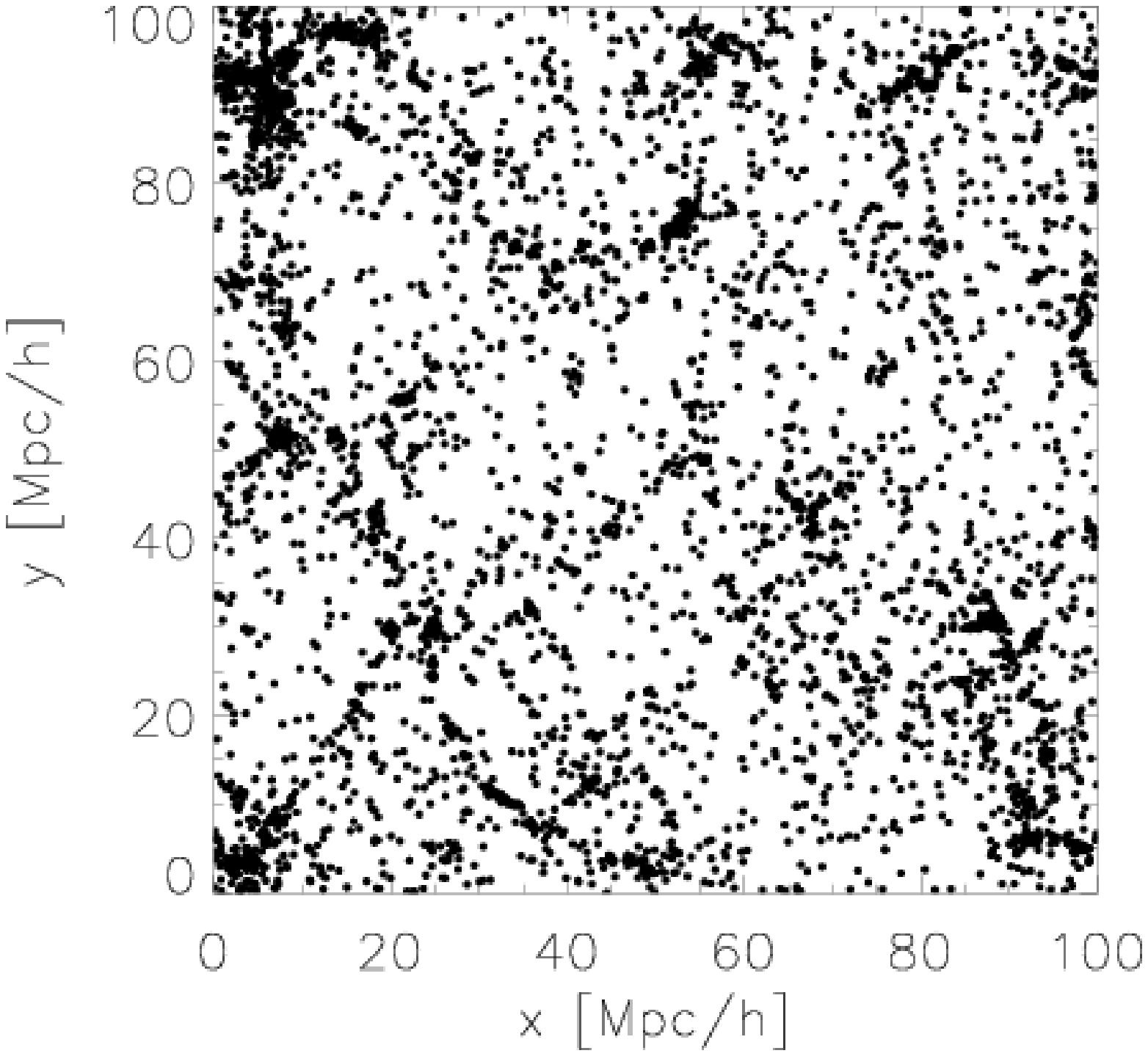}
    \end{minipage}
  \end{center}
 \caption{Upper panel: Three-dimensional representation of the original simulated
          OCDM data set (left) and one representation of a surrogate data set
	  (right). The cube length amounts to 100 Mpc/h. Lower panel: Two-dimensional
	  slice of the original (left) and surrogate (right) data set. All points
	  of a slice of 10 Mpc/h in the middle of the
	  cube (45 Mpc/h $<$ z $<$ 55 Mpc/h) are shown.}
\end{figure*}

Before we show the results of our investigations we want to give
an outlook concerning the techniques of constrained randomisation:
The algorithm described above makes explicit use of the (inverse) Fourier
transformation for evenly binned data sets thus limiting the applicability
of the method. In our case, where we do have evenly binned data and where
we want the surrogates only to have the same power spectrum and the same
amplitudes as the original in configuration space, the method is well suited.
This might not always be the case.  There exist more general approached
for the constrained randomisation of data sets
(Schreiber 1998 and Schreiber \& Schmitz 2000), which rely on the
well-known technique of simulated annealing
(Metropolis et al. 1953 and Kirkpatrick et al. 1983). In this formalism
the constraints are imposed as a cost function which is constructed to have
a global minimum when the constraints are exactly fulfilled. With this more
flexible but very CPU-time consuming approach arbitrary constraints can be
implemented - at least in principle. Thus a systematic analysis of the
different statistical measures and their sensitivity to certain constraints
in the data becomes possible. Such analysis is beyond the scope of this work,
but in future work we will focus on implementing more sophisticated constraints
(e.g. higher-order correlations, phase entropy etc.) in the surrogate
data sets in order to systematically assess statistical
measures used in large scale structure analysis.\\

\section{Results}
With the method described in the previous section we generated a set of 20
three-dimensional surrogate data sets from the original OCDM data.
In Fig. 1 the three-dimensional point distributions as well as a 2-dimensional
slice for the OCDM data and one surrogate realisation are displayed.
Looking at the point distributions one does not see very pronounced
morphological differences between the original and surrogate data set. One can
clearly detect some salient features (e.g. clusters) in both point distributions.
If surrogates are generated applying only a simple phase randomisation without
taking care of the amplitude distribution (see e.g. the example in Chiang 2001)
one obtains a more or less featureless image. Therefore we can conclude that
the additional constraint of preserving the amplitude distribution
in configuration space is responsible for the existence of morphological
features in the point distribution of the surrogate data set. Nevertheless,
a thorough eye-inspection of the original and surrogate data set
might give the impression that the two point distributions have slightly
different topological features. Clusters can be found in the surrogate data
set as well as in the original data whereas the fine filament structures as well
as the voids are not so pronounced in the surrogates.
\begin{figure*}
  \begin{center}
   \begin{minipage}{8.5cm}
     \epsfxsize=8.5cm
     \epsffile{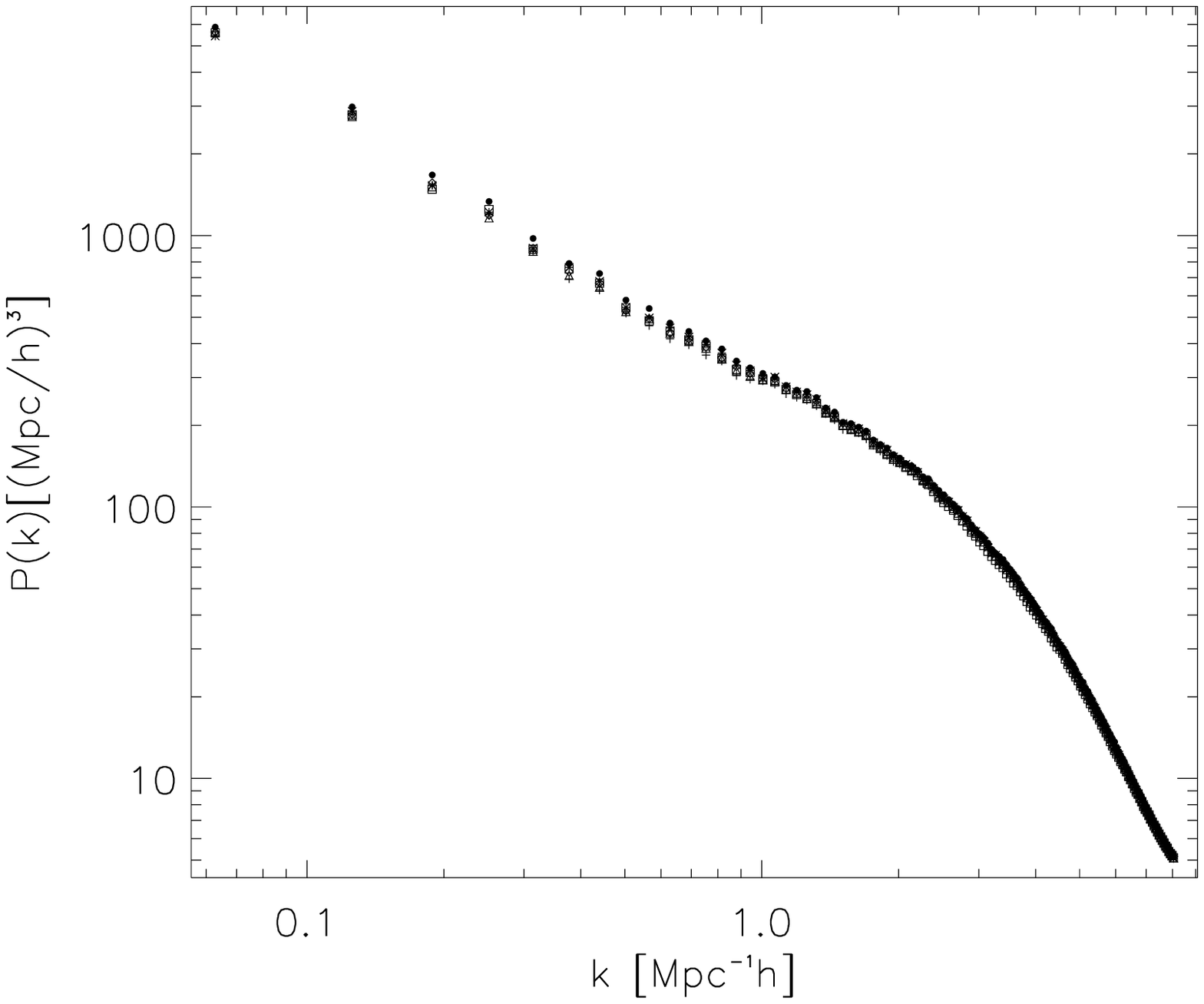}
    \end{minipage}
    \begin{minipage}{8.5cm}
    \epsfxsize=8.5cm
     \epsffile{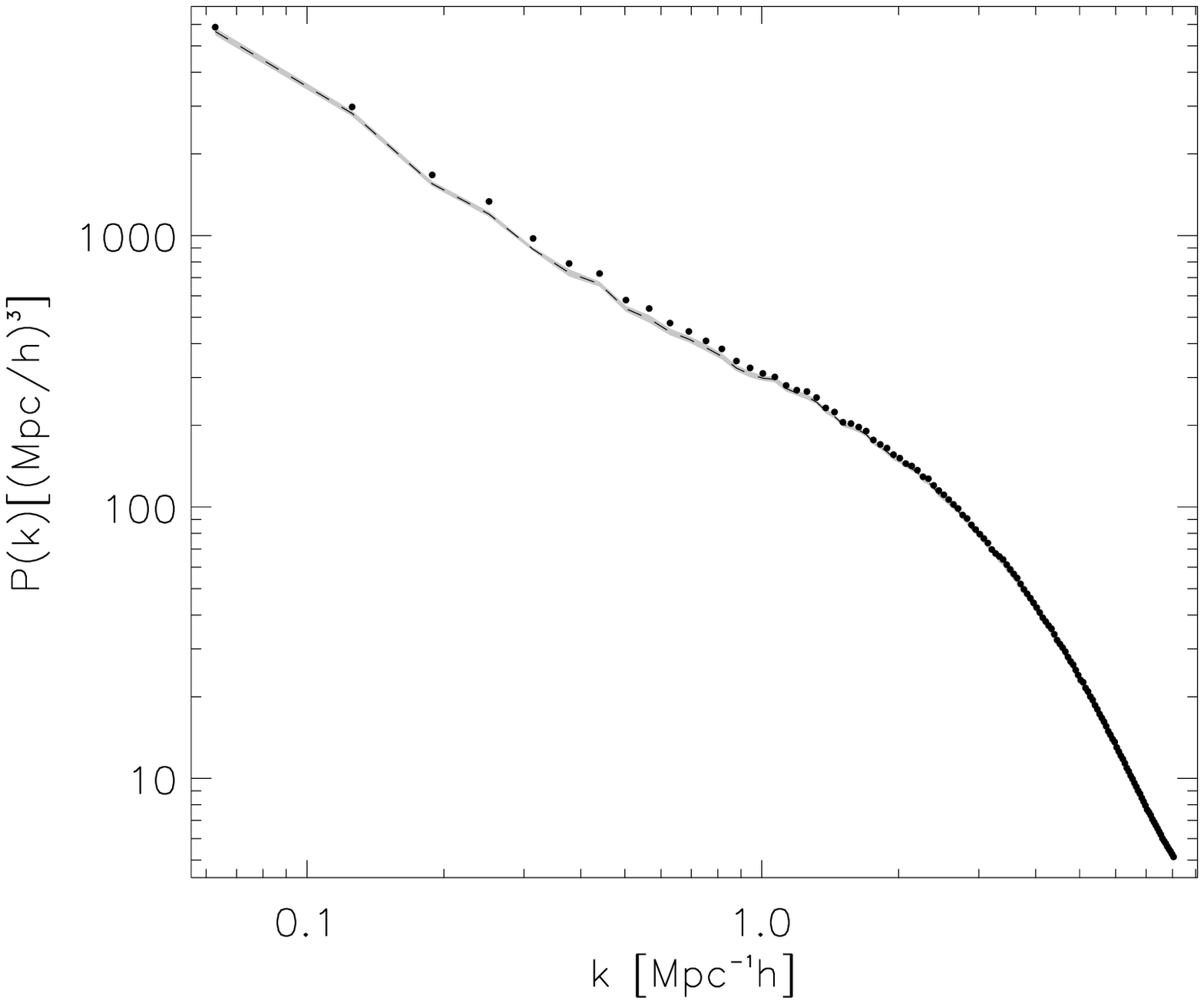}
    \end{minipage}

   \begin{minipage}{8.5cm}
     \epsfxsize=8.5cm
     \epsffile{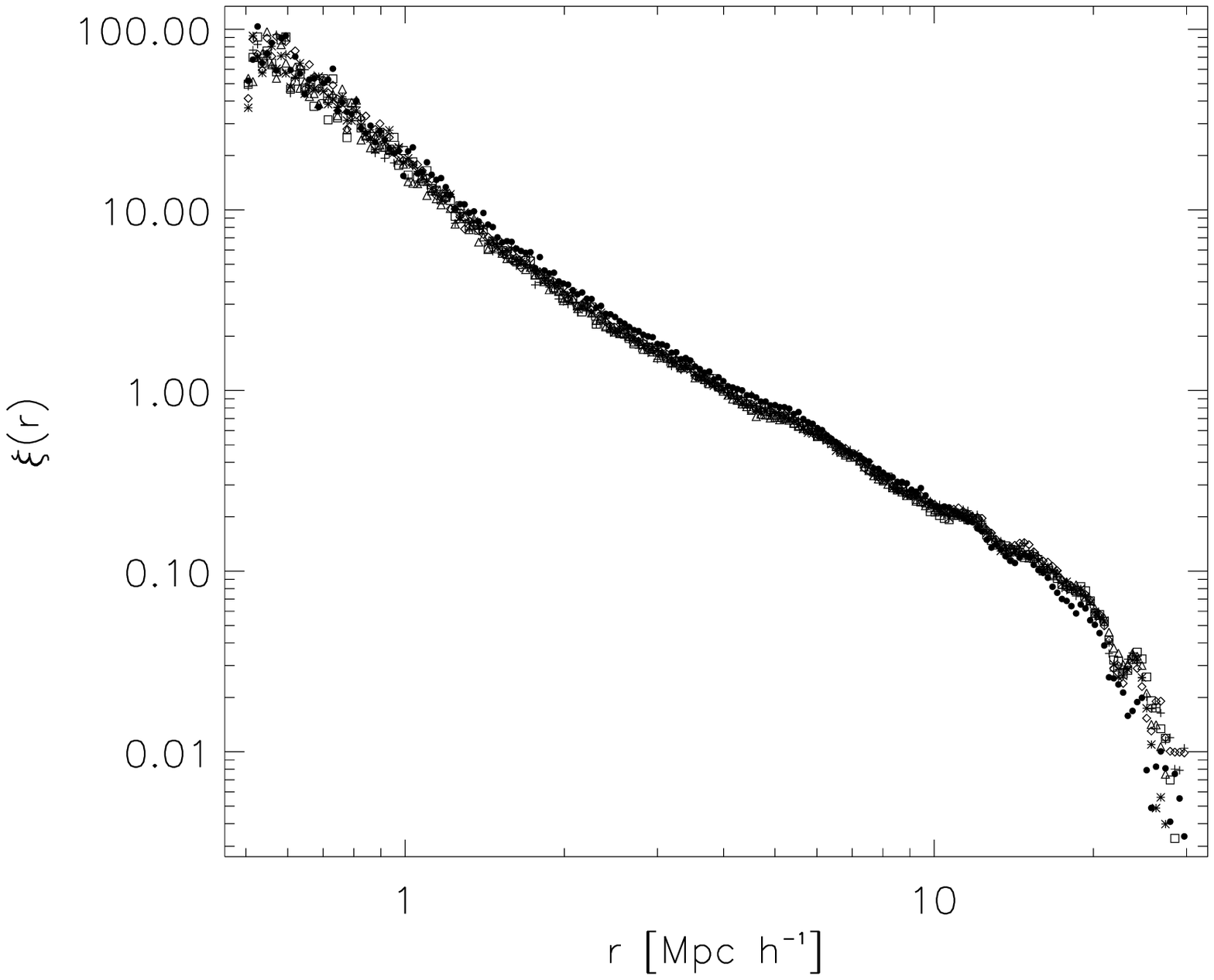}
    \end{minipage}
    \begin{minipage}{8.5cm}
    \epsfxsize=8.5cm
     \epsffile{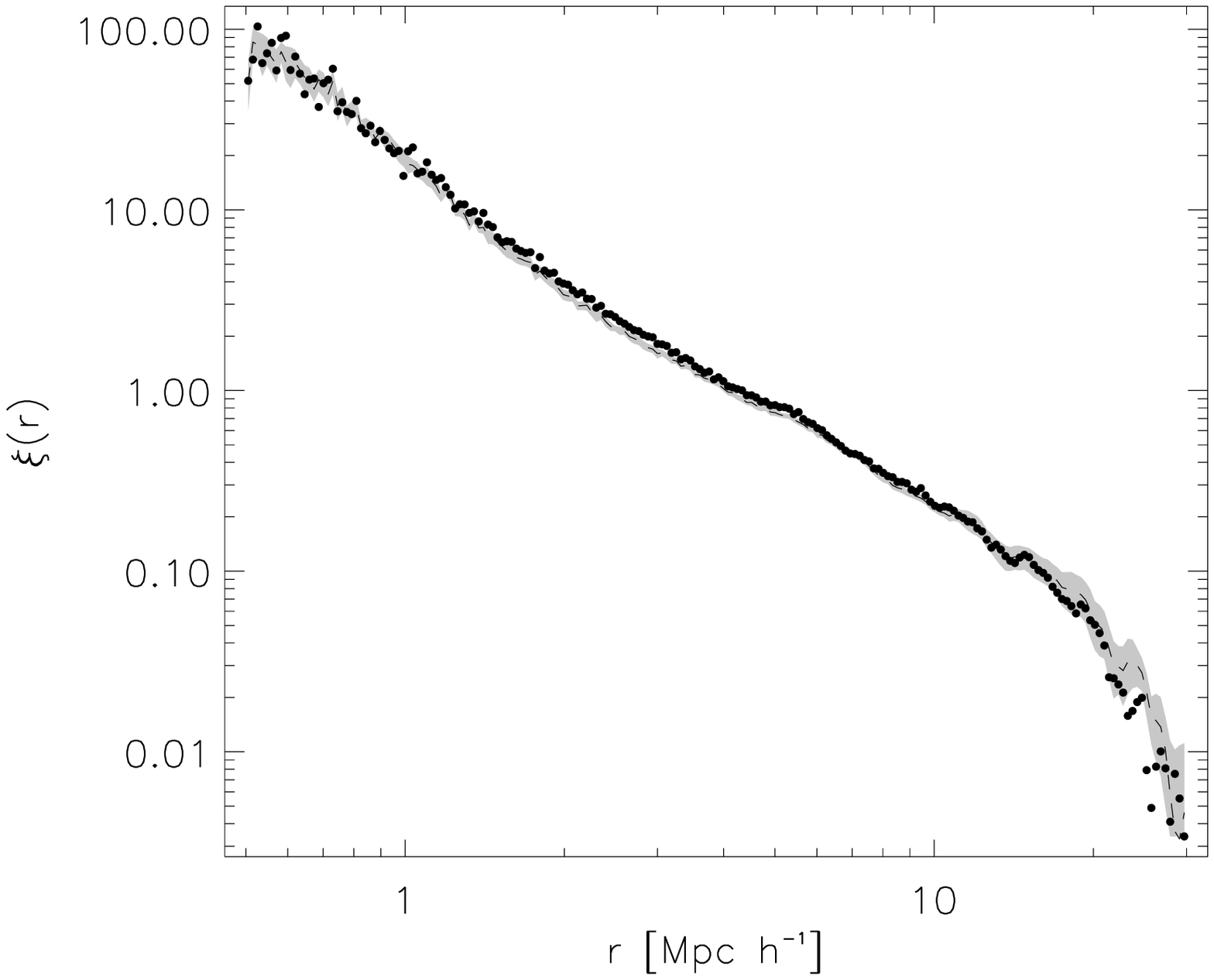}
    \end{minipage}

 \begin{minipage}{8.5cm}
     \epsfxsize=8.5cm
     \epsffile{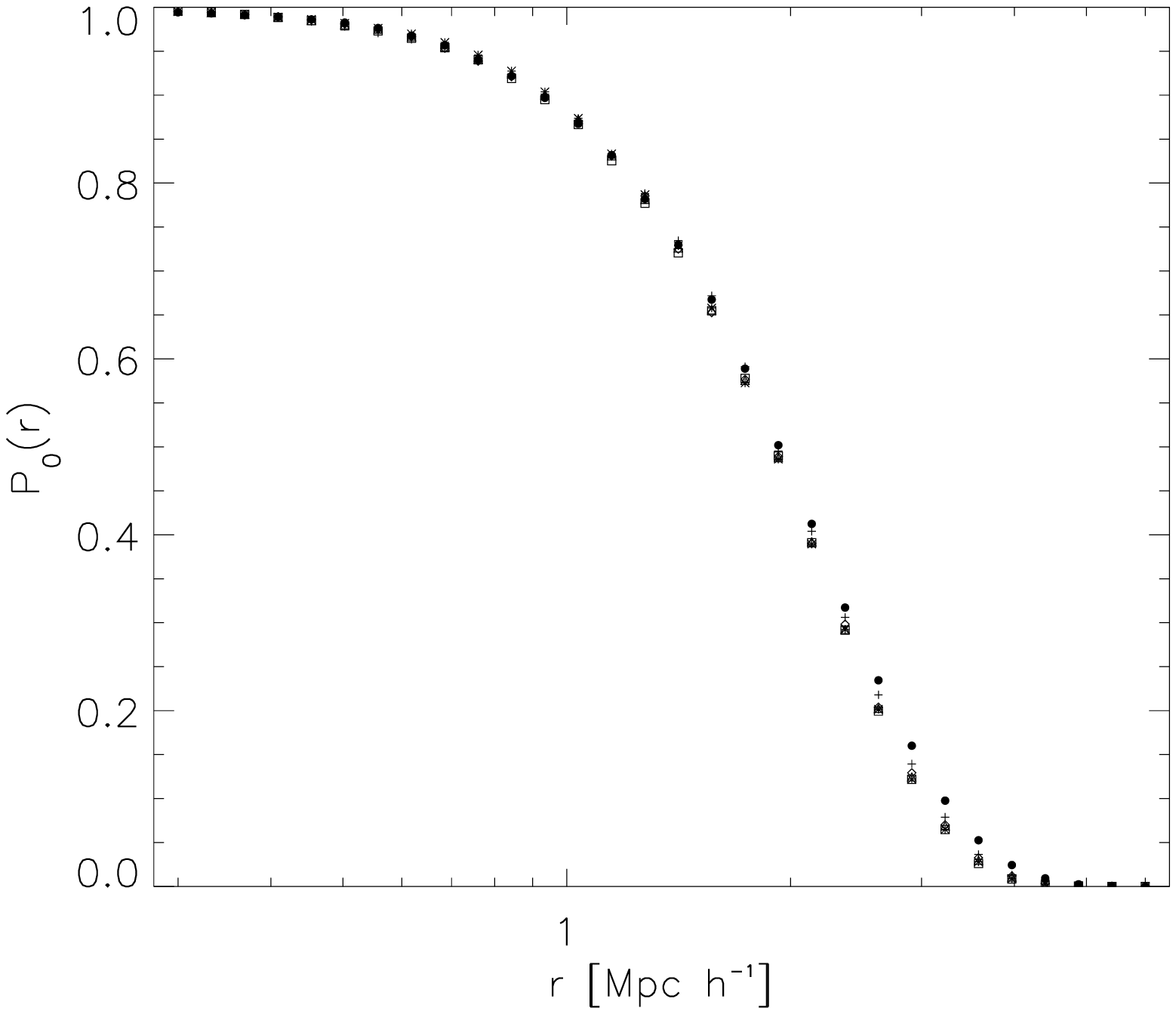}
    \end{minipage}
    \begin{minipage}{8.5cm}
    \epsfxsize=8.5cm
     \epsffile{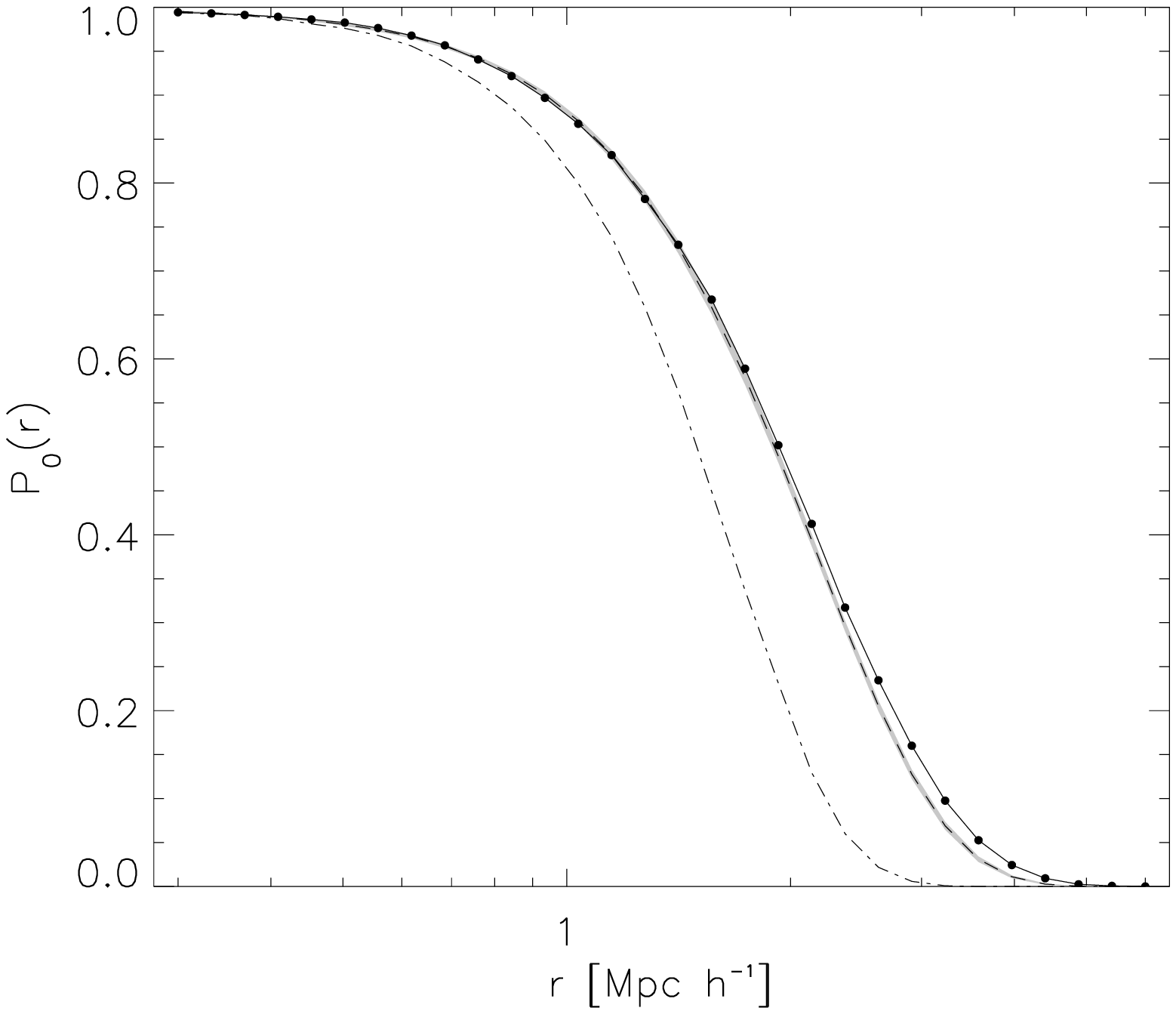}
    \end{minipage}
  \end{center}
  \caption{Power spectrum (upper row), 2-point correlation function (middle row)
           and void probability function (lower row) for the original and
	   surrogate data sets. In the left column the respective measured
	   quantities ($P(k)$, $\xi(r)$ and $P_0(r)$) for the original
	   (filled circles) and five surrogate (other symbols) data sets are shown.
	   In the right column the measured values of the original data sets are
	   compared with the 1 $\sigma$ error region as derived from 20 realisations
	   of the surrogate data. The dash-dotted line in the diagram for the
	   void probability function indicates the graph for $P_0(r)$ in
	   the poissonian case. }
\end{figure*}

The quantitative analysis of the data starts with the calculation of the
power spectrum. In Fig. 2 (upper panel) the power spectra of the
original and surrogate data sets are shown. They are - as required - (almost)
equivalent. For each wave number $k$ the power $P(k)$ for the original
data set lies within or only sightly above the $1 \sigma$- error region
as derived from the power spectra of the 20 surrogates.
Likewise the original data and
surrogates have the same 2-point correlation function as can be seen in
Fig. 2 (middle panel). Only for very low values for $r$ the
2-point correlation function $\xi(r)$ for the original data is outside the
mv phases$1 \sigma$- error region. At these small distances pixelisation effects
become important (pixel size: $0.4$ Mpc/h) so that these deviations
from the expected values are understood.
The void probability function $P_0(r)$ for the data sets is shown in Fig. 2
(lower panel). It can be seen that for higher values of $r$
($r \approx 1.5$ Mpc/h) the VPF for the surrogates is shifted towards the
pure poissonian case and therefore yields slightly lower values than
for the original data which are significantly above
the $1 \sigma$- error area in this
region. Hence a discrimination between the original data and surrogates
using this measure seems possible. The more poissonian-like behaviour
of the surrogate VPF indicates that the higher order correlations in the data
are affected by generating the surrogates, making them more randomly distributed.
However, the VPF for the surrogates still differs significantly from the
pure random case and lies much closer to the VPF of the original data than of
the poisson case.
\begin{figure}
  \begin{center}
     \epsfxsize=8cm
     \epsffile{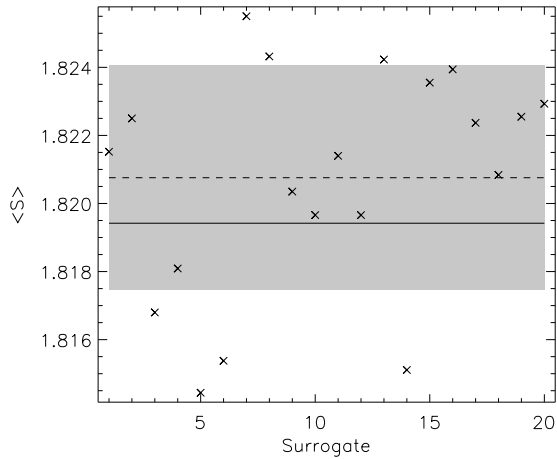}
  \end{center}
 \caption{Mean phase entropy $<S> = (S_x+S_y+S_z)/3$ for the surrogate data
          (crosses) and the original data (solid line). The gray region
	  indicates the 1 $\sigma$ deviation around the mean (dashed line)
	  as derived from the 20 realisations of the surrogates.}
\end{figure}
In order to analyse how the Fourier phases are influenced by randomising the
original data set we calculated the mean phase entropy $<S>$  for all data
sets. The results are displayed in Fig. 3. One can see that for both the
original and surrogate data $< S >$ is significantly lower
than $\ln{2\pi} \approx 1.838$ which is a clear indication that the
phases are not completely uncorrelated but at least partially coupled.
Both classes of data are therefore nongaussian. However, in our sample
the original data set can not be told apart from the surrogates using the
phase entropy. $< S >$ for the OCDM data lies clearly within the $1 \sigma$- error
region close to the mean of $< S >$ for the surrogates. Thus the phase entropy
is a good scalar measure for testing for non-gaussianity but it is
obviously not so well suited to discriminate between different non-gaussian
point distributions. It may be necessary to define more subtle measures
in order to extract the information about the morphology of the structures
contained in the correlations of the Fourier phases more efficiently.\\
\begin{figure*}
  \begin{center}
   \begin{minipage}{8.5cm}
     \epsfxsize=8.5cm
     \epsffile{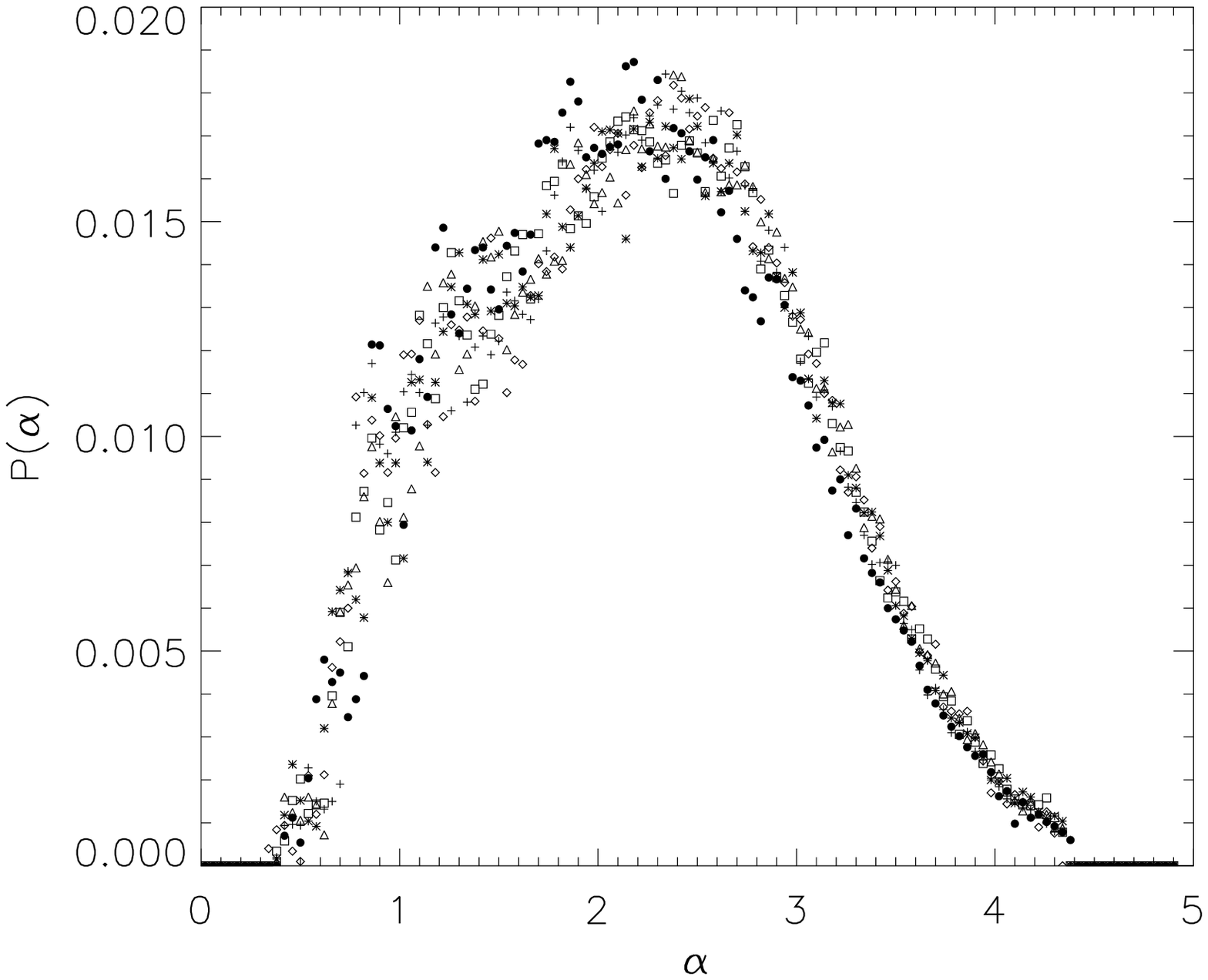}
    \end{minipage}
    \begin{minipage}{8.5cm}
    \epsfxsize=8.5cm
     \epsffile{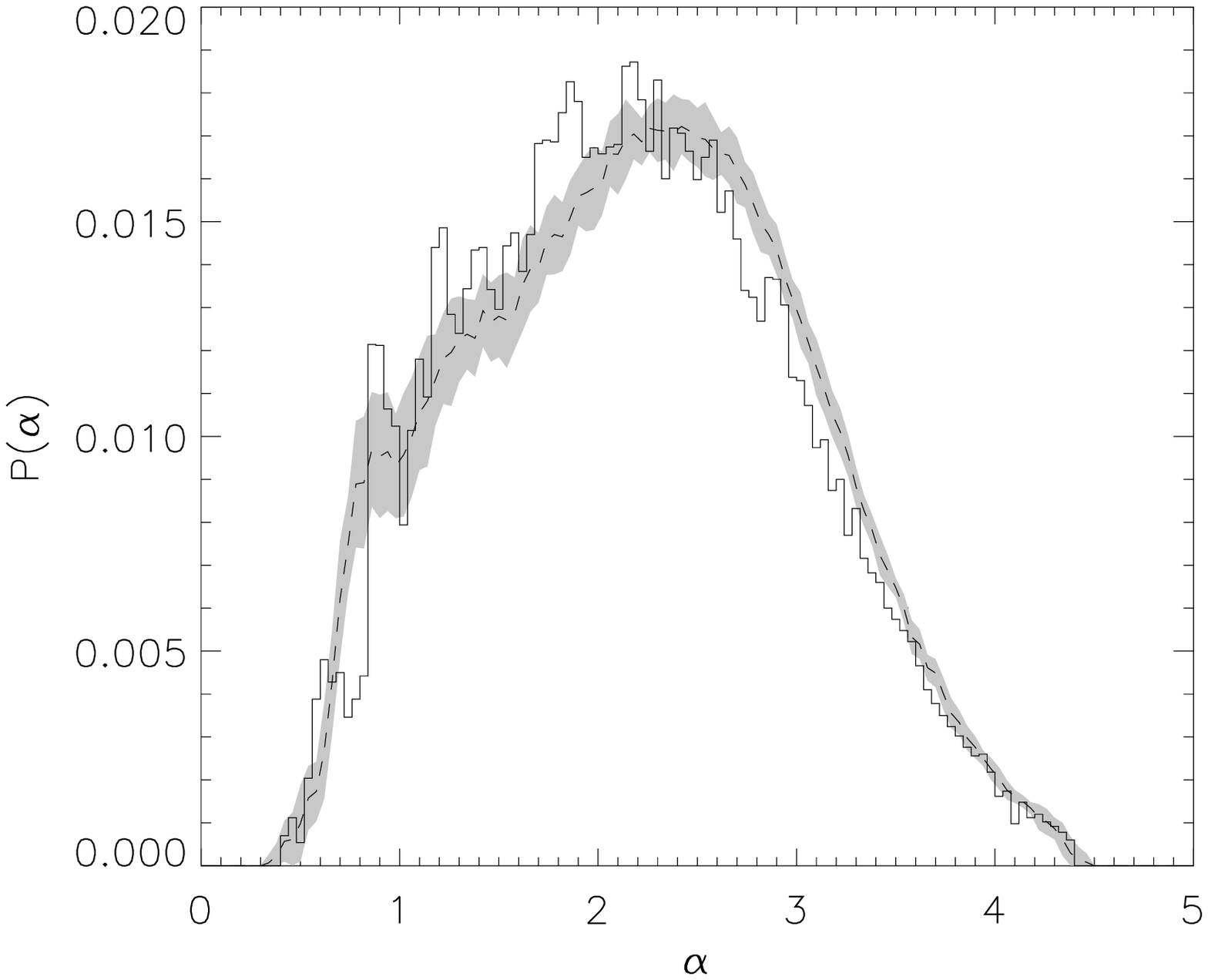}
    \end{minipage}
  \end{center}is
   \caption{Probability density of the scaling indices
            $P(\alpha)$ (r = 4 Mpc/h) for the original and surrogate data
           sets. Left: $P(\alpha)$ for the original (filled circles)
	   and five surrogate (other symbols) data.
	   Right: $P(\alpha)$ for the original data (black histogram)
	   compared with
	   the 1 $\sigma$ error region (gray area) as derived from 20 realisations
	   of the surrogate data.}
\end{figure*}

\begin{figure*}
  \begin{center}
     \epsfxsize=17cm
     \epsffile{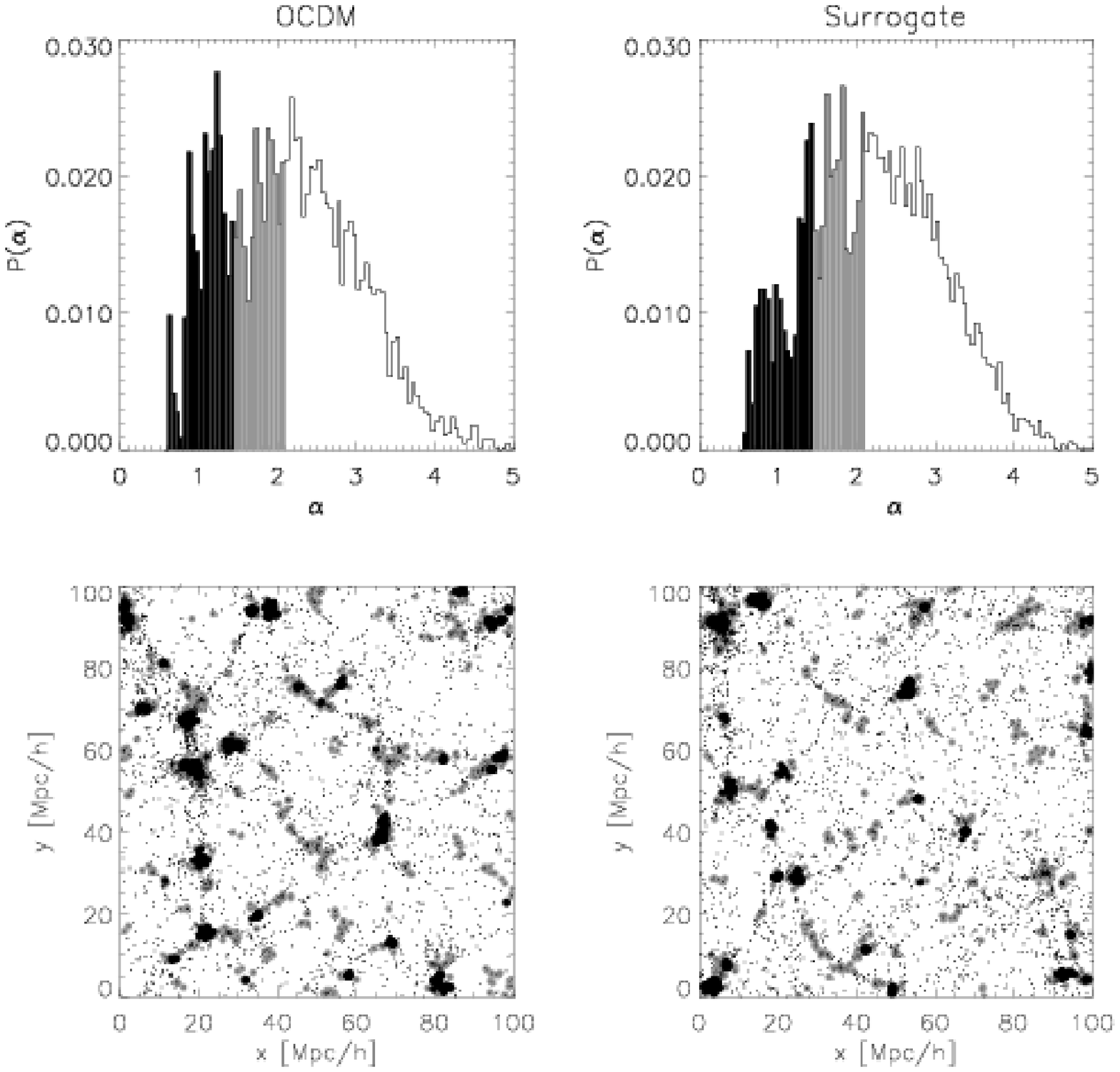}
  \end{center}
  \caption{Probability density $P(\alpha)$ (upper row) and  slices (lower row)
   for the original and one surrogate data set. Points with $\alpha < 1.45$
  are merked in black, points having $1.45 < \alpha < 2.1$ are marked in gray.}
\end{figure*}

The probability distribution $P(\alpha)$ of the weighted scaling indices
$\alpha$ for the original data and surrogates is shown in Fig. 4.
From the comparison between the mean distribution for the surrogates
with its $1 \sigma$-error and the $P(\alpha)$-spectrum for the original
data as displayed in Fig. 4, one can derive several important results.
For a wide range of $\alpha$-vaues the probability distribution $P(\alpha)$
for the OCDM data is significantly outside the $1 \sigma$- error region
of the surrogates. A clear distinction between the surrogates and the
original data is made possible using weighted scaling indices. Thus
the weighted scaling indices and their probability distribution
$P(\alpha)$ have the highest discriminative power of all statistical measures
discussed in this study, making this measure a very promising new candidate
for a refined analysis of the large scale structure.
Furthermore, a more
detailed analysis of the $P(\alpha)$ spectra reveals valuable information
about the morphological differences between the original and surrogate
data sets. For the surrogates the peak in the $P(\alpha)$ distribution
for the OCDM model at $\alpha \approx 0.9$ vanishes, indicating that
the percentage of points belonging to highly overdense cluster-like regions
diminishes. The location of the maximum of the distribution is shifted
from $\alpha \approx 2.0$ for the OCDM data to $\alpha \approx 2.6$
for the surrogates while the height of the peak is retained.
These differences of the $P(\alpha)$ distribution are interpreted as a
loss of wall-like and filament-like structures in
the surrogates with a correspondingly higher percentage of randomly
distributed points, which yields higher values for $P(\alpha)$ in the range
$2.8 < \alpha < 3.5$. In order to visualize the loss of
cluster-like and filament-like structural elements  in the surrogates we
extracted all points in slices of the thickness 10 Mpc/h
(45 Mpc/h $<$ z $<$ 55 Mpc/h) for the original data one surrogate data set.
For these points the $P(\alpha)$-spectrum is determined (see Fig. 5).
We now make use of the possibility offered by the scaling index method to
extract specific structural elements from the point distribution
by selecting the respective regions in the $P(\alpha)$ distribution.
For this purpose all points of the slice with $\alpha < 1.45$ are marked in
black while points having $1.45 < \alpha < 2.1$ are marked in gray.
The differences between the original and surrogate data now become obvious
(see Fig. 5). The marked cluster-like (black) and filament-like (gray)
points are close together and well connected in the OCDM data, clearly
assigning the respective structural elements (clusters or filaments)
to which the points belong to, whereas for the surrogate data
the selected points are more randomly distributed over the plane
and only a smaller percentage of the selected points can clearly
be assigned to certain structural elements.

\section{Conclusions}
We adapted and used the method of surrogate data to analyse
three-dimensional point distributions. It could be shown that
with the help of the method of iteratively refined surrogates
it is possible to generate data sets which have the same power
spectrum and amplitude distribution in configuration space but
differ significantly with respect to their topological structure.
The existence of these topological differences points to nonlinear
processes in the early evolution of the universe and is likely
to be important cosmologically. Hence nonlinear measures need to be
developed to quantify them - after which the consequences for the
different models have to be discussed. Amongst the standard measures
the void probability function gave relatively small differences between
the original and the surrogate data sets, while the 2-point correlation
function and power spectrum were the same (by construction).\\
These results show that linear global measures like
the 2-point correlation function and power spectrum are only of limited
usefulness for the characterisation of the morphological content
of given point distribution and that their discriminative power is,
therefore, also limited. This is mainly due to the fact that these second
order statistical measures are 'blind' to the distributions of Fourier phases,
which are responsible for the fine details of cosmic structures. We further
analysed the distribution of the phases by calculating the phase entropy and
found that the surrogates cannot be told apart from the original data set using
this measures. Therefore, a more sophisticated analysis of the obviously
inherent correlation in the distribution of the phases is required.\\
We showed that the development of nonlinear morphological descriptors,
which are based on the analysis of the local scaling behaviour of the mass
distribution, can offer new possibilities to refine our statistical methods
so that previously ignored subtle but important features can be both detected
and quantitatively characterised. Using such a measure
(weighted scaling indices) a clear distinction based on the different
topological features between surrogates and the original data set
is possible. In the context of evaluating different statistical measures
used in the analysis of large scale structure the method of constrained
randomisation represents a vital tool with which the quality of the newly
developed measures can be tested systematically. Thus a better
quantitative characterisation of the spatial patterns in the
galaxy distribution becomes possible, improving the interpretation and
our outstanding of the large scale structure in the
universe.

\end{document}